\documentclass[9pt,twocolumn,twoside]{pnas-new}

\templatetype{pnasbriefreport} 

\title{Do not forget interaction: Predicting fatality of COVID-19 patients using logistic regression}

\author[a,1]{Feng Zhou}
\author[b]{Tao Chen} 
\author[c,1]{Baiying Lei}

\affil[a]{Department of Industrial and Manufacturing Systems Engineering, University of Michigan, Dearborn, Michigan 48128, USA}
\affil[b]{School of Information Science and Technology, Fudan University, Shanghai, 200433, China}
\affil[c]{School of Biomedical Engineering, Health Science Center, Shenzhen University, Shenzhen, 518060, China}

\leadauthor{Zhou} 

\authorcontributions{F.Z. designed research; F.Z. collected data; F.Z.,T.C., and B.L. analyzed data; and F.Z., T.C., and B.L. wrote the paper.}
\authordeclaration{The authors declare no competing interest.}
\correspondingauthor{\textsuperscript{1}To whom correspondence should be addressed. E-mail: \texttt{fezhou@umich.edu, leiby@szu.edu.cn}}

\keywords{COVID-19 $|$ fatality prediction $|$ logistic regression $|$interaction$|$ } 

\begin{abstract}
Amid the ongoing COVID-19 pandemic, whether COVID-19 patients with high risks can be recovered or not depends, to a large extent, on how early they will be treated appropriately before irreversible consequences are caused to the patients by the virus. In this research, we reported an explainable, intuitive, and accurate machine learning model based on logistic regression to predict the fatality rate of COVID-19 patients using only three important blood biomarkers, including lactic dehydrogenase, lymphocyte (\%) and high-sensitivity C-reactive protein, and their interactions. We found that when the fatality probability produced by the logistic regression model was over 0.8, the model had the optimal performance in that it was able to predict patient fatalities more than 11.30 days on average with maximally 34.91 days in advance, an accumulative f1-score of 93.76\% and and an accumulative accuracy score of 93.92\%. Such a model can be used to identify COVID-19 patients with high risks with three blood biomarkers and help the medical systems around the world plan critical medical resources amid this pandemic.  
\end{abstract}

\begin{document}

\maketitle
\thispagestyle{firststyle}
\ifthenelse{\boolean{shortarticle}}{\ifthenelse{\boolean{singlecolumn}}{\abscontentformatted}{\abscontent}}{}

\dropcap{T}he COVID-19 outbreak has become a global pandemic with community circulation \cite{bento2020evidence}. It has resulted in over 10 million confirmed cases and over 500 thousands deaths worldwide \cite{johns_hopkins_coronavirus_resource_center_covid-19_nodate} by the end of June 2020. It was reported that the fatality rate of critical cases was as high as 61.5\% and the risks were dramatically high with increasing ages and underlying conditions \cite{huang2020clinical}. The majority of the patients developed common symptoms, such as fever, cough, fatigue, and severe ones developed pneumonia, which can be further deteriorated with respiratory failure \cite{wang2020clinical,chen2020epidemiological}. With large numbers of patients confirmed worldwide without effective treatment, the medical systems are imposed with great pressure with severe shortages of intensive care units and other resources.  

Therefore, it is extremely important to develop accurate yet explainable machine learning models that can predict the fatality rate of COVID-19 patients with important prognostic biomarkers. A previous study identified the top three most important  biomarkers, including  lactic dehydrogenase (LDH), followed by lymphocyte (\%), and high-sensitivity C-reactive protein (hs-CRP), and developed a decision tree model to predict the fatality rate of COVID-19 patients with 90\% accuracy for 10.36 days in advance \cite{yan2020interpretable}. Such a model can play a crucial role to allocate critical resources in hospitals appropriately based on the possible fatality rate of the patients. One flaw of such model is that a decision tree can only result in a binary decision that either tell one patient would die or survive without any room in between. In order to provide both accuracy and explainability of the machine learning model with probabilities in predicting the fatality of COVID-19 patients, we proposed a logistic regression model with interaction among the three most important prognostic biomarkers. Our results showed that such a model is able to predict patient fatality with 93.92\% accuracy and with an average 11.30 days in advance. The model further suggested that the threshold fatality probability was 0.8, above which the model predicted the patient would die. Our results can provide the caregivers of the COVID-19 patients with critical information in treating them accordingly.    

\begin{table}
\centering
\caption{Fatality prediction performance of logistic regression models in AUC scores (\%) in mean (s.d.) with 100-round five-fold cross-validation.}
\begin{tabular}{ lrr }
Features & Training sets & Validation sets \\
\midrule
1. LDH & 95.78 (0.65) & 95.67 (2.46) \\
2. LDH, lymphocyte & 95.87 (0.74) & 95.69 (2.44)\\
3. LDH, lymphocyte, hs-CRP & 97.10 (0.86) & 96.64 (2.02)\\
4. LDH, lymphocyte, hs-CRP, \\LDH:lymphocyte & 97.31 (0.58) & 97.01 (2.00)\\
5. LDH, lymphocyte, hs-CRP, \\LDH:lymphocyte, LDH:hs-CRP & \textbf{97.64 (0.56)} & \textbf{97.21 (2.01)}\\
6. LDH, lymphocyte, hs-CRP, \\LDH:lymphocyte, LDH:hs-CRP,\\ lymphocyte:hs-CRP & 97.60 (0.63) & 97.06 (2.02)\\
\bottomrule
\end{tabular}

\end{table}
\section*{Results}
\subsection*{Fitting a Logistic Regression Model}
We used the training dataset with 351 patients without missing any values of the three important features (i.e., LDH, lymphocyte, and hs-CRP) to fit a step-wise logistic regression model with all the second-order interaction items. We used the coefficient of determination adjusted $R^2$  \cite{hastie2009elements}, indicating the portion of variance in the dependent variable (i.e., patient outcome) explained by the independent variables, to determine which interaction items to include. We obtained a model with the largest adjusted $R^2 = 0.797$ with all the three variables, i.e., LDH ($P = 0.000$), lymphocyte ($P = 0.000$), and  hs-CRP ($P = 0.000$), and two interaction items, i.e., LDH:lymphocyte ($P = 0.081$) and LDH:hs-CRP ($P = 0.000$).  
\subsection*{Prediction Performance}
Table 1 shows the fatality prediction performance in terms of area under the curve (AUC) \cite{walter2005partial} scores of logistic regression using 100-round five-fold cross-validation. It can be seen that the logistic regression models are able to accurately predict the fatality outcomes of COVID-19 patients with three important biomarkers. As the number of the features increases, the performance also improves before the last interaction item (i.e.,lymphocyte:hs-CRP) was added. This was consistent with the results of fitting a logistic regression model found above. The model with optimal performance had two (product) interaction items, i.e., LDH:lymphocyte and LDH:hs-CRP.

\begin{figure*}[hbt!]
\centering
\includegraphics[width=17.8cm]{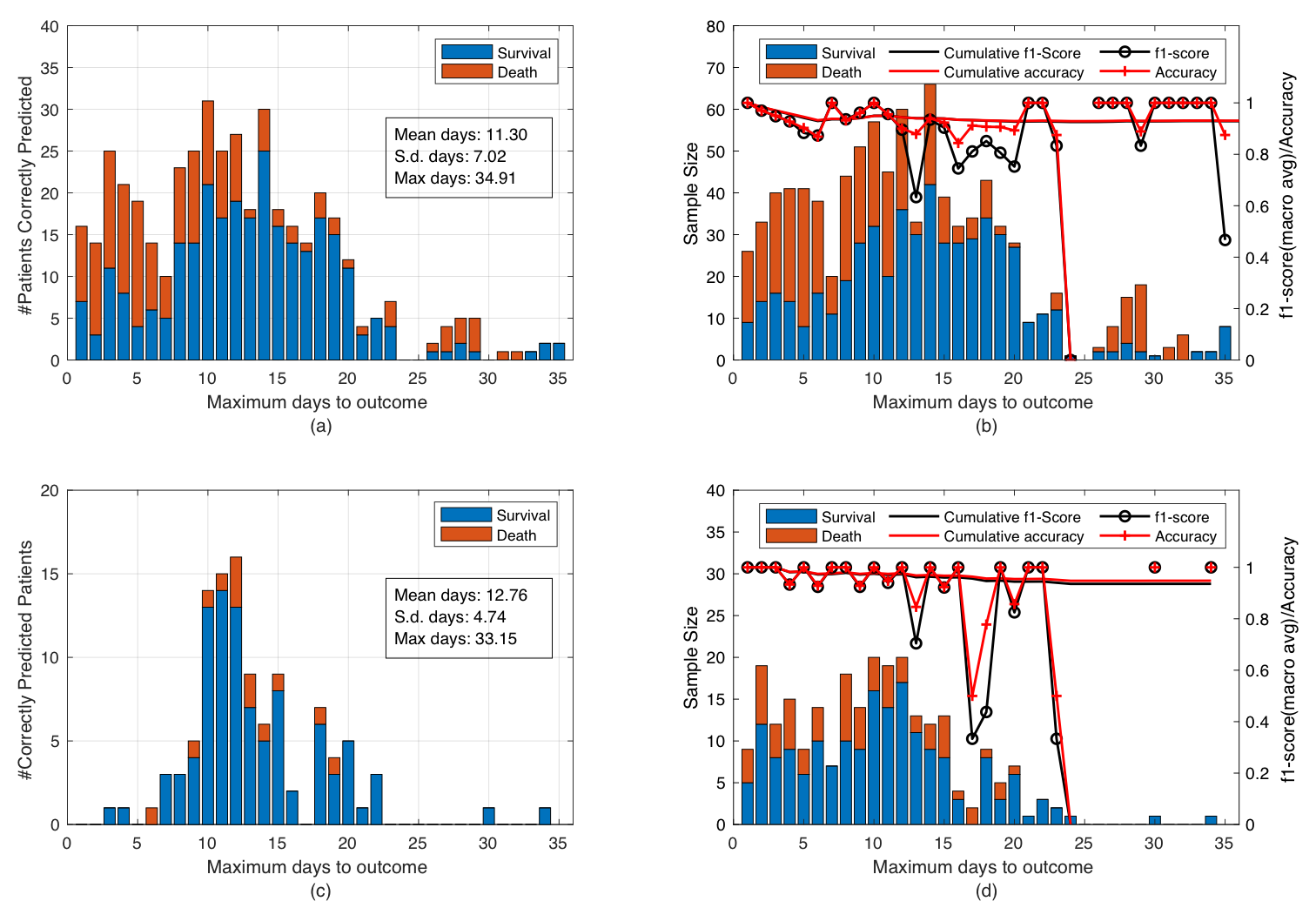}
\caption{The performance of multi-day ahead forecasting of the proposed logistic regression model: Histogram of the maximum days ahead of the correctly predicted patients of (a) all the 485 patients and (c) 110 patients in the external test set; The sample size, f1-score, accuracy in each day and cumulative f1-score and cumulative accuracy over the days of (b) all the the samples and (d) the samples in the external test set. Note there were three negative value records excluded in the figure as they were obtained after the patient was discharged or died.}
\label{fig:frog}
\end{figure*}

\subsection*{An Explainable Logistic Regression Model}
Based on the results in Table 1, we developed an explainable logistic regression model with two interaction items to predict the fatality rate of COVID-19 patients. The modeling of prediction using logistic regression models is transparent and the model can produce fatality probabilities between 0 and 1 rather than a binary value. Using $Y=1$ and $Y=0$ to indicate death and survival, respectively, we can formulate the logistic regression model in the following way \cite{hastie2009elements}:
\begin{equation} \label{eu_eqn}
l\ =\ ln\frac{P( Y=1)}{P( Y=0)} =ln\frac{p}{1-p} =\mathbf{B^{T}X} \ =\sum\limits _{i} \beta _{i} x_{i}, 
\end{equation}
where $\beta_{0}$ is a constant and $\beta_{i}, i=1,..., 5$, are the coefficients of $x_{1}$ (LDH), $x_{2}$ (lymphocyte), $x_{3}$ (hs-CRP), $x_{4}$ (LDH:lymphocyte), and $x_{5}$ (LDH:hs-CRP). We ran 100 rounds of five-fold cross-validation with random search to identify the optimal values of the regularization term $C$ between 0.0001 and 1000 with two types of penalty, i.e., $l1$ and $l2$. This process resulted in 500 sets of the coefficients and we used the median values as the coefficients of the final model in a vector form: $\mathbf{B^{T}} = [-4.976, 1.440e-2, -3.053e-1, 4.378e-2, 4.766e-4, -6.748e-5]$. We also optimized the prediction performance of the model by adjusting the threshold of the death probability both for the external test data and multi-day ahead forecasting below. We found the optimal threshold was 0.8. In other words, when one's fatality probability was larger than 0.8, the model predicted that a patient would die and the model had the best performance.      

The model in Eq.(1) with the identified threshold was then used to predict the outcomes of 110 patients in the external test set that was not used to build the model. Although the data set was rather unbalanced (13 deaths and 97 survivals), the performance of the proposed logistic regression model was promising with 96 true negatives (1 false negative) and 12 true positives (1 false positive). The accuracy, f1-score, and AUC were $98.18\%$, $92.38\%$ and 0.996, respectively.

\subsection*{Multi-Day Ahead Forecasting}
Since there were multiple records of the three biomarkers for each patient, the model was used to forecast patient outcomes multi-days in advance. The samples were obtained by examining the records within each day for each patient. A total number of 909 records was obtained for all the 485 patients with 251 records for the 110 patients in the external test set. For multi-day ahead forecasting, we aimed to obtain the maximum days in advance. For example, if there are $N_{i}$ records in total for patient $i$ and the model was able to predict $m, m<=M$ days ahead, the predicted outcomes and the ground truth of the following maximum $n_{i}, 1<=n_{i}<=N_{i}$ consecutive records need to be the same. Here $m$/$M$ are the days between the dates of the $n_{i}$-th/first records and the date of the final outcome, respectively. For all the records, the model was able to predict 11.30 days (maximum = 34.91) ahead on average with a cumulative f1-score of 93.76\% (see Fig.1 (a) and (b)) and a cumulative accuracy value of 93.92\%. For the 251 records in the external test set, the model was able to predict 12.76 (maximum = 33.15) days ahead on average with a cumulative f1-score of 95.73\% (see Fig.1 (c) and (d)) and a cumulative accuracy value of 96.47\%. Thus, the proposed model can potentially give doctors the time needed to treat the patient accordingly.  

\section*{Discussion}
Built on \cite{yan2020interpretable}, we proposed an explainable, intuitive, and yet accurate prediction model using logistic regression by incorporating two interaction items among the three most important biomarkers, including LDH, lymphocyte (\%) and hs-CRP, which can be easily measured in hospitals. Our model used no extra input information, compared to the decision tree model in \cite{yan2020interpretable}. Unlike the binary decisions produced by the decision tree, the logistic regression model produced a probability of fatality for each patient, which is more consistent with human-friendly explanations of machine learning models \cite{molnar2020interpretable}. For example, one of the rules in the decision tree in \cite{yan2020interpretable} was that IF LDH $> 365 UI^{-1}$, THEN death. Such a binary prediction may be not very intuitive without telling the likelihood of death. As a value of $364 UI^{-1}$ might result in a significantly different fatality probability from a value of $ 64 UI^{-1}$. 

Our model, on the other hand, always gave a probability of death and also identified a threshold probability at 0.8, above which the model predicted that the patient would die. Furthermore, our model also outperformed the decision tree model in terms of average maximum days to the outcome and the cumulative f1-score and accuracy in Fig. 1. 
Such a model can offer the clinicians time to identify high-risk patients before they become critical. Thus, an appropriate treatment strategy for COVID-19 patients depending on their likelihood of death can be made using corresponding medical resources. This can potentially alleviate the shortages of critical medical resources in hospitals in the current situation. 

However, the model was built on a relatively small sample size. More research is needed to further test and optimize the model, taking both explainability and prediction performance into account.

\section*{Materials and Methods}
\subsection*{Samples} The data was originally from \cite{yan2020interpretable}. The model construction was based on the data of 375 (174 died) patients collected between January 10, 2020 and February 18, 2020 from Tongji Hospitals, Wuhan, China. Of them, 24 of the patients had missing data in the three biomarkers and thus were excluded from analysis. The external test data set was collected from another 110 patients (13 died) between February 19, 2020 and February 24, 2020 from the same hospital. We reported the performance of the model using metrics, including AUC \cite{walter2005partial}, micro-ave f1-score  \cite{zhou2011affect}, and accuracy \cite{zhou2020driver}.
\subsection*{Data Analysis}
Interaction items in logistic regression can potentially improve the performance of the model to a great extent \cite{levy2019don}. Hence, we first fitted a logistic regression model using the three most important biomarkers, i.e., LDH, lymphocyte (\%), and hs-CRP identified in \cite{yan2020interpretable} and further identified two interaction items that could be useful to improve the prediction model. Then, we added one item at a time to the logistic regression model with five-fold cross-validation for 100 rounds and verified the two identified interaction items did improve its prediction performance as shown in Table 1. In order to have a model with good generalizability, we used the median values of the coefficients produced from the 500 models when producing the results in Table 1. Finally, this model was used to predict the external test set and multi-day ahead forecasting.

\section*{Conclusions}
The proposed logistic regression model can effectively predict the outcomes of COVID-19 patients with fatality probabilities. The model is accurate, intuitive, and explainable with only three blood biomarkers and two of their interaction items as input, which can potentially help the doctors determine the best treatment route for COVID-19 patients with high risks and optimize the logistic planning in the medical systems around the world amid this COVID-19 pandemic. 

\acknow{We thank the authors in \cite{yan2020interpretable} for
providing their data sets.}

\showacknow{} 

\bibliography{pnas-sample}

\end{document}